\documentstyle[12pt]{article}
\title{An Ultimate Frustration in Classical Lattice-Gas Models} 
\author{Jacek Mi\c{e}kisz \\ Institute of Applied Mathematics \\
and Mechanics \\ Warsaw University  \\ ul. Banacha 2  \\ 02-097
Warsaw, Poland \\ e-mail miekisz@mimuw.edu.pl} 
\begin{document} 
\baselineskip=14pt
\maketitle 
\noindent {\bf Abstract.} We compare tiling systems with square-like tiles
and classical lattice-gas models
with translation-invariant, finite-range interactions between particles.
For a given tiling, there is a natural construction of a corresponding
lattice-gas model. With one-to-one correspondence between particles and tiles,
we simply assign a positive energy to pairs of nearest-neighbor particles
which do not match as tiles; otherwise the energy of interaction is zero.
Such models of interacting particles are called nonfrustrated - all
interactions can attain their minima simultaneously. Ground-state
configurations of these models correspond to tilings; they have
the minimal energy density equal to zero. There are frustrated
lattice-gas models; antiferromagnetic Ising model on the triangular lattice 
is a standard example. However, in all such
models known so far, one could always find a nonfrustrated interaction having
the same ground-state configurations.

Here we constructed an uncountable
family of classical lattice-gas models with unique ground-state measures
which are not uniquely ergodic measures of any tiling system,
or more generally, of any system of finite type.
Therefore, we have shown that the family
of structures which are unique ground states of some translation-invariant,
finite-range interactions is larger than the family of tilings which
form single isomorphism classes. Such ground-state measures cannot
be ground-state measures of any translation-invariant, finite-range,
nonfrustrated potential.

Our ground-state configurations are two-dimensional analogs of
one-dimen\-sional, most homogeneous ground-state configurations
of infinite-range, convex, repulsive interactions in models with
devil's staircases.
\vspace{5mm}


\eject
\section{Introduction}
\newtheorem{theo}{Theorem}
\newtheorem{prop}{Proposition}
We will discuss two families of systems of interacting objects located at 
vertices of the square lattice. A tiling system consists of a finite set
of prototiles, the so-called Wang tiles.
Wang tiles are squares with markings (like notches and dents) on their sides.
These markings define matching rules which tell us which tiles can be
nearest neighbors. Using an infinite number of copies of given prototiles, 
one can tile the plane completely (centers of tiles form the square lattice)
and without overlaps (except boundaries of tiles) such that all matching
rules are satisfied. Naturally, tilings can be seen as structures resulting
from the global maximization of the number of satisfied local matching rules.
It is an outstanding problem to understand why such structures are always
ordered in some sense \cite{rad1}.

A natural generalization of tiling systems are systems of finite
type. Tiling systems are defined by specifying which pairs of tiles cannot be
nearest neighbors. In systems of finite type, we specify which finite patterns
of a fixed bounded size are not allowed.

Our second family consists of two-dimensional classical lattice-gas models.
In such models, sites of the square lattice are occupied by particles
interacting through
translation-invariant, finite-range potentials. Configurations of particles
minimizing the energy density of their interactions are called
ground-state configurations. Like tilings, they are structures optimizing
(minimizing) the sum of local terms. It is an old and still unsolved problem
in solid-state physics, the so-called crystal problem
\cite{born,uhl1,uhl2,bru,and,sim,rad2},
to understand why
ground-state configurations should have a perfect periodic order of crystals
or at least nonperiodic order of recently discovered quasicrystals
\cite{qua,bey}.

For a given tiling system with $n$ prototiles, we can construct the following
lattice-gas model with $n$ types of particles corresponding to tiles.
Two nearest-neighbor particles which do not match as tiles have a positive
interaction energy, say $1$; otherwise the energy of interaction is equal
to zero. Such interactions are obviously nonfrustrated; there are ground-state
configurations minimizing all of them simultaneously. There is a one-to-one
correspondence between such ground-state configurations and tilings
of the plane. In the same manner, a classical lattice-gas
model can be constructed for any system of finite type. Details of this
construction are given in Section 2. 

Here we restrict ourselves to models in which there may be
many tilings or ground-state configurations but there is only one
translation-invariant probability measure supported by them.
Such systems are called uniquely ergodic ones (one may prove that
their unique measures are necessarily ergodic). In case of tilings,
we say that they form a single isomorphism class. In lattice-gas models,
these unique measures are called ground-state measures. They are
zero-temperature limits of translation-invariant Gibbs states describing
an equilibrium behavior of systems of many interacting particles. 

It follows from the above construction that that the family of uniquely
ergodic systems of finite type is contained in the family of uniquely ergodic
classical lattice-gas models with translation-invariant, finite-range
interactions. The main result of this paper is a construction
of an uncountable family of lattice-gas models with finite-range interactions
and with unique ground-state measures. Uncountability is very important here.
There are countably many different bounded patterns of tiles or particles
on a lattice and therefore countably many different finite-type conditions
and hence countably many uniquely ergodic systems of finite type. Our
construction provides us therefore with uncountably many examples of unique
ground-state measures of frustrated, translation-invariant, finite-range
interactions which are not unique ground-state measures of any nonfrustrated,
translation-invariant, finite-range interactions and consequently
they are not uniquely ergodic measures of any system of finite type.
These are measures with an irrational density of different types of particles
and are supported by nonperiodic ground-state configurations.
On the other hand, measures supported by a periodic configuration
and its translates are necessarily of finite type.

Ground-state configurations of our model are two-dimensional analogs
of one-dimensional, most homogeneous configurations present in models
with infinite-range, convex, repulsive interactions \cite{hub}.
Such models exhibit a devil's staircase structure of ground-state measures
\cite{bak,aub,grif}.

In Section 2, we describe systems of finite type and general classical
lattice-gas models with unique ground-state measures. In Section 3, we discuss
a one-dimensional model with a devil's staircase. Section 4 contains our
construction of a classical lattice-gas model with an ultimate frustration.
A short discussion follows in Section 5.

\section{Tilings, systems of finite type, and lattice-gas models}
We begin by discussing {\bf tilings} with square-like tiles.
Our tiles are squares with markings on their sides. These markings
define matching rules which tell us which tiles can be nearest neighbors.
In every tiling, centers of squares form the square lattice ${\bf Z}^{2}$.
Tilings can be therefore represented by assignments of tiles to the sites
of ${\bf Z}^{2}$, i.e., by elements of
$\Omega= \{1, \ldots , n\}^{{\bf Z}^{2}}$, where $n$ is the number of
different types of tiles, the so-called
prototiles. We are interested in uniquely ergodic tiling systems.
In such systems, although there are possibly many tilings, using the same
family of prototiles, there are unique translation-invariant
probability measures on $\Omega$ which are supported by them.
If matching rules
allow only translates of one periodic tiling, then the unique tiling
measure assigns an equal probability to all of these translates.
Generally, a unique tiling measure, $\mu$, gives equal
weights to all tilings and can be obtained as
the limit of averaging over a given tiling $X$
and its translates $\tau_{{\bf a}}X$ by lattice vectors ${\bf
a} \in {\bf Z}^{d}$: $\mu= \lim_{\Lambda \rightarrow {\bf Z}^{d}}
\frac{1}{|\Lambda|} \sum _{{\bf a} \in \Lambda} \delta
(\tau_{{\bf a}} X)$, where $\delta (\tau_{{\bf a}}X)$ is the
probability measure assigning probability $1$ to $\tau_{{\bf a}}X$.
There are examples of tiling systems with unique measures supported
by nonperiodic tilings \cite{rob,moz,gru,sen}.

A natural generalization of tiling systems are systems of finite type.
Let $G$ be a translation-invariant, closed subset of $\Omega$ and $\mu$
a uniquely ergodic, translation-invariant measure supported by $G$.

$(\Omega, G, \mu)$ is a {\bf dynamical system of finite type},
if there exist $C_{i} \in \{1,...,n\}^{\Lambda_{i}}$ for some finite
$\Lambda_{i} \subset {\bf Z}^{d}$ and $i=1,...,m$ such that
$$G=\{X: X(\tau_{{\bf a}}\Lambda_{i}) \neq C_{i} \; \; for \; \; all \; \; 
{\bf a} \in {\bf Z}^{d} \; \; and \; \; any \; \; i=1,...,m\}.$$
In other words, $G$ is defined by the absence of a finite number
of certain local configurations.  

In {\bf classical lattice-gas models}, every site of the ${\bf Z}^{d}$
lattice,
$d\geq 1$, can be occupied by one of $n$ different particles. Configurations
of lattice models are assignments of particles to the lattice sites, i.e.,
elements of $\Omega = \{1,\ldots , n\}^{{\bf Z}^{d}}.$
If $X \in \Omega$ and $\Lambda \subset {\bf Z}^{d}$, then $X(\Lambda)
\in \Omega_{\Lambda} = \{1,\ldots ,n\}^{\Lambda}$ is a projection of
$X$ on $\Lambda.$ Particles interact through generally many-body
potentials. A {\bf potential} $\Phi$ is a family of real-valued
functions, $\Phi_{\Lambda}$ on $\Omega_{\Lambda}$, for all finite
$\Lambda \subset {\bf Z}^{d}.$ If $\Phi_{\Lambda} = 0$ when
$diam(\Lambda)>r$ for a certain $r>0,$ then we say that $\Phi$ has a
finite range $r$. We assume that $\Phi$ is translation invariant, i.e.,
$\Phi_{\Lambda + {\bf a}}(\tau_{{\bf a}}X)=\Phi_{\Lambda}(X)$,
where $\tau_{{\bf a}}$ is the translation by the lattice
vector ${\bf a} \in {\bf Z}^{d}$
and $\Phi_{\Lambda}(X) \equiv \Phi_{\Lambda}(X(\Lambda))$. 
 
For a finite $\Lambda \subset {\bf Z}^{d}$, a {\bf Hamiltonian} of particles
in $\Lambda$ can be written as
$$H^{\Phi}_{\Lambda}= \sum_{V \subset \Lambda} \Phi_{V}.$$

$Y$ is a {\bf local excitation} of $X$, $Y \sim X$, $Y, X \in \Omega$, if
there exists a finite $\Lambda \subset {\bf Z}^{d}$, such that $Y=X$
outside $\Lambda$.

The {\bf relative Hamiltonian} is defined as 
$$H^{\Phi}(Y,X)= \sum_{\Lambda \subset
{\bf Z}^{d}} (\Phi_{\Lambda}(Y)- \Phi_{\Lambda}(X)) \; \; for \;
\; Y \sim X.$$
Observe, that for finite-range potentials, there are only a finite number
of nonzero terms in the above sum.

$X \in \Omega$ is a {\bf ground-state configuration} of a potential $\Phi$ if 
$$H^{\Phi}(Y,X) \geq 0$$
for every $Y \sim X,$ i.e., one cannot lower the energy of a ground-state
configuration by its local change (on a finite subset of lattice sites).

The {\bf energy density} $e(X)$ of a configuration $X$ is defined as
$$e(X)= \liminf_{\Lambda \rightarrow {\bf Z}^{d}}
\frac{H^{\Phi}_{\Lambda}(X)}{|\Lambda|},$$
where $\Lambda \rightarrow {\bf Z}^{d}$ in some certain sense.

One can prove that if $X$ is a ground-state configuration, then $X$ has
the minimal energy density, i.e., $e(X) \leq e(Y)$ for every $Y \in \Omega$.
It means that local conditions contained in the definition of a
ground-state configuration force the global minimization of the energy
density.

Although, for any given Hamiltonian, the set of ground-state configurations
is nonempty, it may not contain any periodic configuration
\cite{rad3,rad4,mier1,jac1}. 

In our models, there is a unique translation-invariant probability
measure on $\Omega$, supported by ground-state configurations. 
It is then necessarily the zero-temperature limit of equilibrium states
(translation-invariant Gibbs states). We call it the {\bf ground state}
of a given model.
 
A potential $\Phi$, for which there exists a configuration minimizing
simultaneously all interactions $\Phi_{\Lambda}$, is called nonfrustrated.
Such a configuration is of course a ground-state configuration.

Formally, a potential $\Phi$ is nonfrustrated or is called 
an m-potential \cite{hol,sla},
if there exists
a configuration $X \in \Omega$ such that
$$\Phi_{\Lambda}(X)=min_{Y}\Phi_{\Lambda}(Y)$$
for every finite $\Lambda \subset {\bf Z}^{d}.$

\begin{theo}
There is one-to-one correspondence between dynamical systems of finite type
and uniquely ergodic ground-state measures of classical lattice-gas models
with nonfrustrated, transla\-tion-invariant, finite-range potentials. 
\end{theo}
{\bf Proof:} Let $(\Omega, G, \mu)$ be a dynamical system of finite type
defined by the absence of $C_{i}, i=1,...,m.$ We define
a translation-invariant potential $\Phi$ such that $\Phi_{\Lambda}(X(\Lambda))
=1$, if $\Lambda$ is a translate of $\Lambda_{i}$ for some $i$ and
$X(\Lambda)=C_{i}$, and zero otherwise. $\Phi$ is obviously nonfrustrated and
$\mu$ is the unique ground state of $\Phi$.

Conversely, let $\Phi$ be a nonfrustrated, translation-invariant,
finite-range potential with a unique ground-state measure $\mu$
supported by a set $G$
of ground-state configurations. Let $X \in \Omega$ be such that
$$ \Phi_{\Lambda}(X)= min_{Y}\Phi_{\Lambda}(Y) \; \; for \; \; any
\; \; finite \; \; \Lambda.$$
$G$ is then defined by the absence of local configurations $X(\Lambda)$
such that $\Phi_{\Lambda}(X(\Lambda)) \neq min_{Y}\Phi_{\Lambda}(Y).$
Hence $(\Omega, G, \mu)$ is a dynamical system of finite type. $\Box$
\vspace{2mm}

The goal of this paper is to construct a classical lattice-gas model
with a frustrated, translation-invariant, finite-range potential
and with a uniquely ergodic ground-state measure $\mu$ which is not
a uniquely ergodic measure of any dynamical system of finite type or
equivalently not a ground-state measure of any nonfrustrated,
translation-invariant, finite-range potential. 

\section{One-dimensional devil's staircases and the most homogeneous
configurations}

One of the examples of a frustrated potential is provided
by the following lattice-gas model with infinite-range
interactions \cite{hub}. Every site of the one-dimensional
lattice ${\bf Z}$ can be occuppied by one particle or be empty.
Particles at a distance $n$ interact through a convex, repulsive potential
$V_{n}$: $V_{n}>0,$
$V_{n+1}+V_{n-1} \geq 2V_{n}$ for $n>1$, and $V_{n} \rightarrow 0$
as $n \rightarrow \infty$. For any given density $\rho$ of particles, one
can find the energy density $e(\rho)$ of ground-state configurations
\cite{hub}.
For any rational $\rho$, there is a unique (up to translations) periodic 
ground-state configuration with that density of particles. It has the
following
property. Let $x_{i} \in {\bf Z}$ be a coordinate of the $i$th particle.
Then there exists a sequence of natural numbers $d_{j}$ such that 
$x_{i+j}-x_{i} \in \{d_{j}, d_{j}+1\}$ for every $i \in {\bf Z}$
and $j \in {\bf N}$. Configurations with such property are called the
{\bf most homogeneous configurations}. 

Of course, if we do not fix the density, particles want to be as far
one from another as possible, so the vacuum is the only ground state.
Now we introduce a chemical potential $h>0$ and pass to the grand-canonical
ensemble. Particles are now frustrated -
they do not want to be on the lattice because of the interactions between them
and at the same time they want to be on the lattice because of
the chemical potential.
To find the energy density of a ground state we have to minimize
\begin{equation}
f(\rho)=e(\rho)-h\rho.
\end{equation}  
Now, $e(\rho)$ is differentiable at every irrational $\rho$ and is 
nondifferentiable at any rational $\rho$ \cite{aub}. However,
as a convex function,
it has a left derivative $d^{-}e(\rho)/d\rho$ and a right derivative 
$d^{+}e(\rho)/d\rho$ at every $\rho$. It follows that to have a ground state
with an irrational density, $\rho$, of particles, one has to fix
$h(\rho)=de(\rho)/d\rho$. For any rational $\rho$, one has the interval
of chemical potentials $h \in [d^{-}e(\rho)/d\rho, d^{+}e(\rho)/d\rho]$.
One can show that the sum of lengths of these intervals has the length
of the interval of all considered values of chemical potentials. 
We have obtained a {\bf complete devil's staircase} \cite{bak,aub}.

As we have already mentioned, for any rational $\rho$, there is a unique
(up to translations) periodic ground-state configuration with that density
of particles - there is a unique ground-state measure. For any irrational
$\rho$, there are uncountably many ground-state configurations which are
the most homogeneous configurations. Now we will show that there is still
the unique ground-state measure supported by them. 
\begin{prop}
For any $0 \leq \rho \leq 1$, there exists a unique sequence $d_{n}$
such that the corresponding most homogeneous configurations have $\rho$
as their density of particles.
\end{prop}
{\bf Proof:} Let $\rho^{n}(d_{n})$ be the density of pairs of particles
which are the $n$th neighbors at a distance $d_{n}$ in the most
homogeneous configurations. The following system
of equations have unique solutions for $d_{n}$ and $0 \leq \rho^{n}(d_{n}),
\rho^{n}(d_{n}+1) \leq 1$, for any $n \geq 1$:
\begin{equation}
\rho^{n}(d_{n})+\rho^{n}(d_{n}+1)=\rho,
\end{equation}
\begin{equation}
d_{n}\rho^{n}(d_{n})+(d_{n}+1)\rho^{n}(d_{n}+1)=n.
\end{equation}
\begin{theo}
For any $0 \leq \rho \leq 1$, there is a unique translation-invariant
probability measure (the ground-state measure of the corresponding
Hamiltonian) supported by the most homogeneous configurations such that $\rho$
is their density of particles.
\end{theo}
{\bf Proof by the induction:} Assume that there two such measures, $\mu_{1}$
and $\mu_{2}$. Denote by $\mu_{1}(d_{1})$ the density, in $\mu_{1}$, of pairs
of two successive particles at a distance $d_{1}$, by $\mu_{1}(d_{1},d_{1}+1)$
the density of triples of three successive particles with succesive
distances $d_{1}$ and $d_{1}+1$, and generally,
by $\mu_{1}(P_{n})$ with $P_{n}=(p_{1},...,p_{n}), p_{i}
\in \{d_{1},d_{1}+1\},i=1,...,n$, the density of $(n+1)th$ tuples of
$n+1$ successive
particles with $p_{i}$ as successive distances between them.
Analogously, we introduce densities for $\mu_{2}.$
We will show that $\mu_{1}(P_{n})=\mu_{2}(P_{n})$
for every $P_{n}$ and every $n \geq 1$.
We will use the induction on $n$.

The above equality for $n=1$ follows
from the fact that both $\mu_{1}$ and $\mu_{2}$ have the same density of
particles.

Let $n=2$. If $P_{2}=(d_{1},d_{1})$, then let $P'_{2}=(d_{1},d_{1}+1)$. Then
\begin{equation}
\mu_{i}(P'_{2})= \mu_{i}(d_{1}+1), \; \; i=1,2
\end{equation}
and therefore
\begin{equation}
\mu_{1}(P'_{2})= \mu_{2}(P'_{2}).
\end{equation}
We have
\begin{equation}
\mu_{1}(P_{2})+\mu_{1}(P'_{2})= \mu_{2}(P_{2})+\mu_{2}(P'_{2})
\end{equation}
and therefore
\begin{equation}
\mu_{1}(P_{2})=\mu_{2}(P_{2}).
\end{equation}
All three remaining types of $P_{2}$ can be treated in an analogous way.

Now assume the equality for any $P_{k}$ with fixed $k \geq 2$.
If $P_{k+1}$ is of the form $(d_{1},P_{k-1},d_{1})$ for some $P_{k-1}$
and $P'_{k+1}=(d_{1}+1,P_{k-1},d_{1})$, then
\begin{equation}
\mu_{1}(P'_{k+1}) = \mu_{1}(d_{1}+1,P_{k-1}).
\end{equation}
\begin{equation}
\mu_{2}(P'_{k+1}) = \mu_{2}(d_{1}+1,P_{k-1}).
\end{equation}
By the induction assumption the right-hand sides of (8) and (9) are equal
and hence the left-hand sides of (8) and (9) are equal.
Now again by the induction assumption we have
\begin{equation}
\mu_{1}(P_{k-1},d_{1})=\mu_{2}(P_{k-1},d_{1})
\end{equation}
and hence
\begin{equation}
\mu_{1}(P_{k+1})+\mu_{1}(P'_{k+1})=\mu_{2}(P_{k+1})+\mu_{2}(P'_{k+1}).
\end{equation}
It follows that
\begin{equation}
\mu_{1}(P_{k+1})=\mu_{2}(P_{k+1}).
\end{equation}
All three remaining types of $P_{k+1}$ can be treated
in an analogous way.  $\Box$
\vspace{2mm}

To summarize, for every chemical potential, there is a unique
ground-state measure of the corresponding Hamiltonian. Therefore, there are
uncountably many Hamiltonians with unique strictly ergodic
ground-state measures.

One of the goals of this paper is to investigate if one can obtain similar
results in two-dimensional models with strictly finite-range interactions.
Let us mention at this point, that for any finite-range
interaction in one dimension, there exists at least one periodic
ground-state configuration \cite{radsch,mier2}. Hence a devil's staircase
cannot appear in one-dimensional classical lattice gas models with
finite-range, translation-invariant interactions.
\section{A model with an ultimate frustration}

Let us first describe particles of our model. They
correspond to square tiles with diagonal, horizontal, and
vertical markings. There is a tile without any markings
and there are tiles with one or two diagonal markings as shown 
in Fig.1.  A tile with the horizontal, vertical, and two diagonal
markings is called a cross and is shown in Fig.2.
All other tiles are called arms and are shown in Fig.3.

Our first finite-type condition is a nearest-neighbor
or a next-nearest-neighbor matching rule which says
that a line of markings cannot end.
This is translated into a nearest-neighbor or a next-nearest-neighbor
interaction between
two particles in the standard way. Two nearest-neighbor
or next-nearest-neighbor
particles which do not match as tiles have a positive interaction energy,
$J_{2}>0$; otherwise the energy is equal to zero.

Our second finite-type condition allows only certain patterns of five
vertically or horizontally successive tiles. Namely, 
among five vertically succesive tiles there must be at least one
arm with the horizontal marking or a cross and there cannot
be two such tiles at a distance smaller than four. Analogously,
among five horizontally successive tiles there must be
at least one arm with the vertical marking or a cross and there
cannot be two such tiles at a distance smaller than four. Again,
this is translated into a five-body interaction by simply assigning
a positive energy, $J_{5}>0$, to all forbidden patterns;
allowed five-particle patterns have zero energy. 

Finally, we have a three-site condition which forces every arm with
diagonal markings to have a cross as one of its nearest neighbors.
A respective coupling constant is denoted by $J_{3}>0$.

A {\bf broken bond} is a local configuration of particles which
does not satisfy a finite-site condition.

Now we will construct ground-state configurations of a lattice-gas model
with the above finite-range translation-invariant interactions.
Looking just at horizontal and vertical markings
we see an infinite grid of infinite horizontal
and vertical lines such that nearest-neighbor parallel lines
are at a distance four or five. These are 
the only configurations of particles corresponding to tilings which satisfy
the two-site and five-site conditions described above. Now we will show
that the three-site condition forces distances between lines to follow
the rule (discussed in Ch.3) of the most homogeneous configurations of atoms
on the one-dimensional lattice ${\bf Z}$ \cite{hub}.

\begin{prop}
Let $X$ be a configuration which satisfies the two-site and five-site
conditions.
Let $x_{i}$ be a double-sided sequence of $x$ coordinates of vertical lines
and $y_{j}$ be a double-sided sequence of $y$ coordinates of horizontal
lines in $X$. Then $X$ satisfies the three-site condition (and therefore
it is a ground-state configuration) if and only if there is
a sequence of natural numbers $d_{n}$ such that for every $n\geq 1$ either
\begin {equation}
x_{i+n}-x_{i}, y_{j+n}-y_{j} \in \{d_{n}, d_{n}+1\}
\end{equation}
or
\begin{equation}
x_{i+n}-x_{i}=d_{n}, y_{j+n}-y_{j} \in \{d_{n}-1,d_{n}, d_{n}+1\}
\end{equation}
or
\begin{equation}
x_{i+n}-x_{i} \in \{d_{n}-1,d_{n}, d_{n}+1\}, y_{j+n}-y_{j}=d_{n}
\end{equation}
for every $i$ and $j$.
\end{prop}
{\bf Proof by the induction:} The five-site condition forces (13) to be
satisfied with $d_{1}=4$.
Now let us consider lines which are next-nearest neighbors. Let us 
assume, without loss of generality, that $x_{i+2}-x_{i}=10$ and
$y_{j+2}-y_{j}=8$.
A diagonal line passing through a lattice site $(x_{i},y_{j})$ intersects
a horizontal
line at a lattice site $(x_{i+8},y_{j+8})$ which violates the three-site
condition.
Conversely, if condition (13) is satisfied with $d_{2}=8$ or $d_{2}=9$, 
or (14) or (15) with $d=9$, then any
diagonal line passing through a lattice site $(x_{i},y_{j})$ intersects
nearest and next-nearest horizontal and vertical lines at a distance
at most one from a cross.  

We will proceed now with the second step of the induction. The following
statement is assumed to be true: a diagonal line passing through
a lattice site $(x_{i},y_{j})$ intersects $k$ nearest horizontal
and vertical lines
at a distance at most one from a cross, if and only if, for every $n=1,...,k$,
(13) or (14) or (15) is satisfied for every $i$ and $j$. Now we have
to show that this statement is true for $k+1.$ Let us assume, without loss
of generality, that $x_{i+k}-x_{i}=d_{k}+1$ and $y_{j+k}-y_{j}=d_{k}.$
If $x_{i+k+1}-x_{i+k}=5$ and $y_{j+k+1}-y_{j+k}=4$, so none of the above 
conditions are satisfied, then the diagonal line intersects a vertical line at
a lattice site $(x_{i+k+1},y_{j+k+1}+2)$ and a horizontal line at 
a lattice site $(x_{i+k+1}-2,y_{j+k+1})$, so the three-site condition
is violated. In all three remaining cases, (13) or (14) or (15) is satisfied
and intersections are at a distance at most one from a cross. $\Box$    
\vspace{2mm}

Observe, that if at least for one $n$, (14) or (15)
is satisfied for every $i$ and $j$, then $X$ is periodic,
with a period $d_{n}$ in $x$ or $y$ direction respectively.
The density of arms is therefore rational; in fact it is equal to $2n/d_{n}$.
Let us note that our model has
ground-state configurations with all possible densities of horizontal
and vertical markings (counted together) satisfying following inequalities:
$2/5 \leq \rho_{m} \leq 1/2.$ Therefore, it has
uncountably many different ground-state measures. On the other hand,
if one fixes the irrational density of horizontal and vertical markings,
then our model has a unique ground-state measure which we denote
by $\mu_{\rho_{m}}$. For any rational $\rho_{m}$,
we have many ground-state measures. In both cases we have that
the density of crosses, $\rho_{cr}=(\rho_{m}/2)^{2}$. 
Now we introduce chemical potentials, $h_{cr} < 0$ for crosses and
$h_{a} > 0$ for arms. For fixed $\rho_{m}$, the energy density of
any configuration satisfying all finite-site conditions is given
by a convex function
\begin{equation}
f(\rho_{m})= -(h_{cr}-2h_{a})(\rho_{m}/2)^{2}-h_{a}\rho_{m}.
\end{equation}
Minimization of $f$ with respect to $\rho_{m}$ gives us
\begin{equation}
\rho_{m}=\frac{2h_{a}}{2h_{a}-h_{cr}}.
\end{equation} 
Now we will show that when the density of horizontal and vertical markings,
$\rho_{m}$, is fixed, then $\mu_{\rho_{m}}$ is the only ground-state measure
of the Hamiltonian including all finite-site conditions and chemical
potentials, and its energy density is given by (16) and (17).

\begin{prop}
If $J_{5}$ is sufficiently big, then the density of broken five-site bonds
is equal to zero in any ground-state measure.
\end{prop}
{\bf Proof:} If among five vertically (horizontally) successive particles
in a configuration $X$ there are not any particles with the horizontal
(vertical) marking
or there are particles with the horizontal (vertical) marking
at a distance smaller than four,
then we either put there a particle with the horizontal (vertical) marking or 
remove a particle with the horizontal (vertical) marking.
In may happen that we have to put or
remove nearby some particles with the horizontal (vertical) marking, in order
not to create other broken five-site bonds. During this process we may create
some broken two-site or three-site bonds. However, if $J_{5}$ is
sufficiently big, the above procedure decreases the energy
and therefore the configuration $X$ is not a ground-state configuration. 
$\Box$
\vspace{2mm}

Now we will show that also the density of broken two-site and three-site bonds
is zero in any ground-state measure with a fixed $\rho_{m}$.
Let $\rho$ be a density of broken bonds in a probability measure $\mu$
which has zero density of broken five-site bonds.
Let $n=2^{m}$ be such that $1/n^{2}<\rho$. Let $S= \{{\bf a} \in {\bf Z}^{2}:
0 \leq a_{1}, a_{2} <n\}$. We call $\tau_{{\bf b}}S$, ${\bf b} \in
{\bf Z}^{2}$, 
an {\bf r-square} of a configuration $X$ in the support of $\mu$, if
the number of vertical
markings, $n_{v}$, and the number of horizontal markings, $n_{h}$, satisfy
the following inequalities:
\begin{equation}
(r-1)n < |n_{v}-n_{h}| \leq rn
\end{equation} 
for a natural number $r>1$ and
\begin{equation}
0 \leq |n_{v}-n_{h}| \leq n
\end{equation}
for $r=1$.
\begin{prop}
If $S$ is an r-square of $X$, $r > 1$,then the number of broken bonds,
$B$, in X(S) is bigger than $r^{2}/9.$ 
\end{prop} 
{\bf Proof:} If $r=2$, then it follows from Proposition 2 that there is
a broken bond in $X(S)$. If $r>2$, then we divide $S$ into four squares of
the size $n/2$. If a smaller square is a 2-square (with $n/2$ in (18)),
then there is a broken
bond in it. We call such a square a {\bf good} square. If a smaller
square is a 1-square (with $n/2$ in (19)), then we call it a {\bf bad}
square. Every r-square with $r>2$ we divide again into four squares.
We continue this procedure untill all squares are either good or bad squares.
Let $D=\sum_{i}\frac{r_{i}}{2^{k_{i}}}$, where the summation
is with respect to all
good and bad squares; $r_{i}=2$ for every good square, $r_{i}=1$
for every bad one and $k_{i}$ is the number of divisions to get a given
square. We have that $D \geq r$. Let $G$ be the number of good squares.
Proposition 2 tells us that $B \geq G$. Now we have to prove that $G>D^{2}/9$.

The above division procedure can be represented by a hierarchical directed
tree with vertices corresponding to squares and edges joining a square
with its four subsquares. Good and bad squares are final vertices
of such tree. Among four final squares connected to a common square,
there must be at least one good square. Let us notice that when we enlarge
a tree by connecting a good square to three bad squares and one good
square, we increase $D$ and leave $G$ unchanged. Therefore, to prove
the above bound, it is sufficient to consider such trees that all good
squares are of the same size and no square is connected to more than one good
square. Let $k$ be the smallest number such that all squares of size
$n/2^{k}$ which are not final ones, are connected to three bad squares
and one 
good square. We may also assume that there are no bad squares of sizes
bigger than $n/2^{k}$. Otherwise, we could
take a part of a tree connected to a square of size $n/2^{k}$ (changing
this square to a bad one) and connect it to a bad square of size $n/2^{k'}$
with $k'< k$, increasing in this way $D$ and not changing $G$. 
Let us assume now that there are $s$ bad squares of size $n/2^{k}$;
$0 \leq s < 3\times 4^{k-1}$. For such a tree   
\begin{equation}
G=4^{k}-s,
\end{equation}
\begin{equation}
D < s/2^{k}+3(4^{k}-s)/2^{k}.
\end{equation}
It follows from (20) and (21) that
\begin{equation}
G>D^{2}/9
\end{equation}
hence the induction step is finished.

The equality in (22) is attained
in the infinite tree with $k=s=0$. $\Box$

\begin{theo}
For a fixed density of horizontal and vertical markings, $\rho_{m}$,
for the Hamiltonian specified by chemical potentials
$h_{cr}$ and $h_{a}$ and all finite-type conditions described above,
$\mu_{\rho_{m}}$ is the only ground-state measure.
\end{theo}
{\bf Proof:} If the density of horizontal markings, $\rho_{hm}$, is equal
to the density of vertical markings, $\rho_{vm}$, then in the absence
of broken horizontal and vertical lines (broken two-site,
nearest-neighbor bonds), $\rho_{cr}= (\rho_{m}/2)^{2}$.
We may decrease the density of crosses but for every removed cross
we have to create a broken horizontal or vertical line and this increases
the energy if $J_{2}$ is sufficiently big.
It shows that in the case 
of $\rho_{hm}=\rho_{vm}$, $\mu_{\rho_{m}}$ is the only ground state. Let
us suppose now that $\rho_{hm} \neq \rho_{vm}$,
$\rho_{hm}= \rho_{m}/2 - \alpha$ and $\rho_{vm}= \rho_{m}/2 + \alpha$,
$\alpha >0$. Again, let us assume first that there are no broken horizontal
and vertical lines. Then
\begin{equation}
\rho_{cr}= (\rho_{m}/2)^{2} - \alpha^{2}.
\end{equation} 
Denote by $\rho_{r}$ the density of $r$-squares and by $\rho$ the density
of broken bonds in a configuration $X$. We have
\begin{equation}
\alpha = (\rho_{vm}-\rho_{hm})/2 \leq \frac{1}{2n^{2}} \sum_{r}\rho_{r}rn
\end{equation}
and by the Jensen's inequality we obtain
\begin{equation}
\alpha^{2} \leq \sum_{r}\rho_{r}\frac{r^{2}}{4n^{2}}.
\end{equation}
Hence, at most we may decrease the density of crosses by the amount on the
right-hand side of (25), so if $J_{2},J_{3}> (10/4)|h_{cr}|$, then $\rho=0$,
if $X$ is a ground-state configuration. We had to put $10$ instead of $9$
in the bound in order to deal with 1-squares by using $1/n^{2} < \rho$.
Of course, we may decrease farther the density of crosses, but as before, for
every removed cross we have to create a broken line.  $\Box$ 
\vspace{2mm}

To summarize, for a fixed chemical potential $h_{a}$, for every irrational
density of horizontal and vertical markings, $2/5 \leq \rho_{m} \leq 1/2$,
it follows from (16) and (17) that there exists
a chemical potential $h_{cr}$ given by
\begin{equation}
h_{cr}= 2h_{a}(1-1/\rho_{m})
\end{equation}
such that the corresponding Hamiltonian has a unique ground-state measure,
$\mu_{\rho_{m}}$, with
$\rho_{m}$ as the density of horizontal and vertical markings. 
Therefore, there are uncountably many uniquely ergodic ground-state measures
on a phase diagram of our model.

\section{Conclusions}

Two potentials are called {\bf equivalent} if they have the same
relative Hamiltonians and therefore the same ground states and Gibbs states.
 
In \cite{jac2} we constructed a model with a frustrated,
translation-invariant,
nearest-neighbor potential for which there does not exist an equivalent,
nonfrustrated, translation-invariant, finite-range potential. An important
feature of that model is the absence of periodic ground-state
configurations. It was a first deterministic lattice-gas model in which
a global minimum of energy is not a sum of local (in space) minima.
To be more precise, one cannot minimize the energy of interacting particles
by minimizing their energy in a finite volume and all its translates,
no matter how big is the volume. In fact, if you take a finite box af any size
and find a configuration of particles in this box which minimizes
the energy of their interactions, then such a configuration,
called a local ground-state configuration, cannot be a part of an
infinite-lattice ground-state configuration (compare also \cite{pen}).

Here we constructed models with unique nonperiodic ground states
which are not unique ground states of any nonfrustrated potential,
equivalent or not. It means that our nonperiodic ground-state configurations
are represented by tilings without any local matching rules.
Such situation was investigated in microscopic models of quasicrystals.
It was suggested in \cite{gahl} that some quasiperiodic structures
from a single isomorphism class (a uniquely ergodic ground state
in our terminology) do not allow for any local matching rules
but can be stabilized by some local cluster interactions.   

To summarize, we have shown that the family of structures which are
unique ground states of some translation-invariant, finite-range
interactions is larger than the family of tilings which form single
isomorphism classes.
\vspace{3mm}

\noindent {\bf Acknowledgments.} I would like to thank the Polish Committee
for Scientific Research, for a financial support under the grant
KBN 2P03A01511.

\eject
\begin{picture}(800,100)(-70,0)

\put(-100,100){\line(1,0){80}}
\put(-100,20){\line(1,0){80}}
\put(-100,20){\line(0,1){80}}
\put(-20,20){\line(0,1){80}}

\put(40,100){\line(1,0){80}}
\put(40,20){\line(1,0){80}}
\put(40,20){\line(0,1){80}}
\put(120,20){\line(0,1){80}}
\put(40,20){\line(1,1){80}}

\put(180,100){\line(1,0){80}}
\put(180,20){\line(1,0){80}}
\put(180,20){\line(0,1){80}}
\put(260,20){\line(0,1){80}}
\put(180,100){\line(1,-1){80}}

\put(320,100){\line(1,0){80}}
\put(320,20){\line(1,0){80}}
\put(320,20){\line(0,1){80}}
\put(400,20){\line(0,1){80}}
\put(320,20){\line(1,1){80}}
\put(320,100){\line(1,-1){80}}

\end{picture}

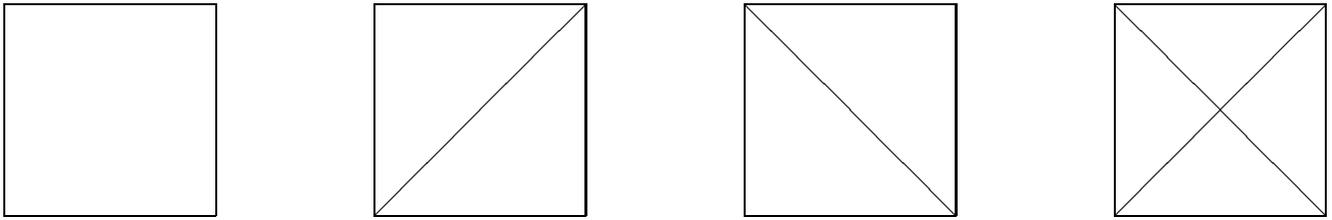
Fig.1. Tiles without horizontal and vertical markings 

\vspace{15mm}

\begin{picture}(400,100)(-70,0)
\put(-100,100){\line(1,0){80}}
\put(-100,20){\line(1,0){80}}
\put(-100,20){\line(0,1){80}}
\put(-20,20){\line(0,1){80}}
\put(-100,20){\line(1,1){80}}
\put(-100,100){\line(1,-1){80}}
\put(-100,60){\line(1,0){80}}
\put(-60,20){\line(0,1){80}}
\end{picture}

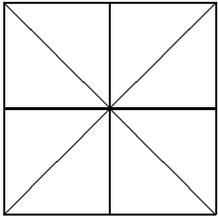
Fig.2. A cross

\begin{picture}(800,300)(-70,50)

\put(-100,300){\line(1,0){80}}
\put(-100,220){\line(1,0){80}}
\put(-100,220){\line(0,1){80}}
\put(-20,220){\line(0,1){80}}
\put(-100,260){\line(1,0){80}}

\put(40,300){\line(1,0){80}}
\put(40,220){\line(1,0){80}}
\put(40,220){\line(0,1){80}}
\put(120,220){\line(0,1){80}}
\put(80,220){\line(0,1){80}}

\put(180,300){\line(1,0){80}}
\put(180,220){\line(1,0){80}}
\put(180,220){\line(0,1){80}}
\put(260,220){\line(0,1){80}}
\put(180,300){\line(1,-1){80}}
\put(180,260){\line(1,0){80}}
\put(180,220){\line(1,1){80}} 

\put(320,300){\line(1,0){80}}
\put(320,220){\line(1,0){80}}
\put(320,220){\line(0,1){80}}
\put(400,220){\line(0,1){80}}
\put(320,220){\line(1,1){80}}
\put(320,300){\line(1,-1){80}}
\put(360,220){\line(0,1){80}}

\put(-100,160){\line(1,0){80}}
\put(-100,80){\line(1,0){80}}
\put(-100,80){\line(0,1){80}}
\put(-20,80){\line(0,1){80}}
\put(-100,80){\line(1,1){80}}
\put(-100,120){\line(1,0){80}}

\put(40,160){\line(1,0){80}}
\put(40,80){\line(1,0){80}}
\put(40,80){\line(0,1){80}}
\put(120,80){\line(0,1){80}}
\put(40,160){\line(1,-1){80}}
\put(40,120){\line(1,0){80}}

\put(180,160){\line(1,0){80}}
\put(180,80){\line(1,0){80}}
\put(180,80){\line(0,1){80}}
\put(260,80){\line(0,1){80}}
\put(180,80){\line(1,1){80}}
\put(220,80){\line(0,1){80}}

\put(320,160){\line(1,0){80}}
\put(320,80){\line(1,0){80}}
\put(320,80){\line(0,1){80}}
\put(400,80){\line(0,1){80}}
\put(320,160){\line(1,-1){80}}
\put(360,80){\line(0,1){80}}

\end{picture}

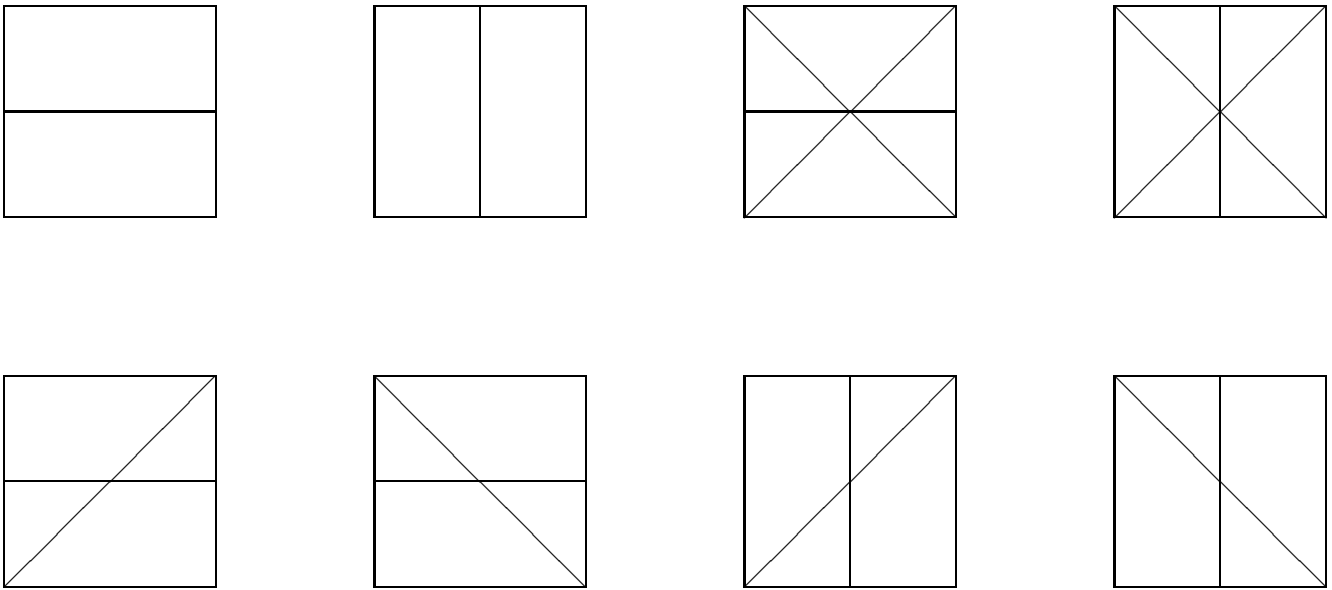
Fig.3. Arms

\end{document}